\documentclass{nature}
\usepackage{graphicx}
\makeatletter
\let\saved@includegraphics\includegraphics
\AtBeginDocument{\let\includegraphics\saved@includegraphics}

\makeatother
\usepackage{dcolumn}
\usepackage{bm}
\usepackage{color,soul}
\definecolor{purple}{RGB}{127,0, 255}

\usepackage{caption}
\usepackage{ragged2e}
\usepackage{amssymb}
\usepackage{float}
\usepackage{amsmath}
\newcommand{\tc}{$T_\textrm{c}$ }

\newcommand{\sr}[1]{}

\title{Coupled Ferroelectricity and Superconductivity in Bilayer $T_d$-MoTe$_2$}

\author{Apoorv Jindal$^{1}$, Amartyajyoti Saha$^{2,3}$, Zizhong Li$^{4}$, Takashi Taniguchi$^{5}$, Kenji Watanabe$^{5}$, James C. Hone$^{6}$, Turan Birol$^{3}$, Rafael M. Fernandes$^{2}$, Cory R. Dean$^{1}$, Abhay N. Pasupathy$^{1,\dagger}$, Daniel A. Rhodes$^{4,\dagger}$}

\begin{document}

\maketitle

\begin{affiliations}
 \item Department of Physics, Columbia University, New York, NY, USA
 \item School of Physics and Astronomy, University of Minnesota, Minneapolis, MN, USA
  \item Department of Chemical Engineering and Materials Science, University of Minnesota, Minneapolis, MN, USA
 \item Department of Materials Science and Engineering, University of Wisconsin, Madison, WI, USA
 \item National Institute for Materials Science, Tsukuba, Japan
 \item Department of Mechanical Engineering, Columbia University, New York, NY, USA
 \item[$^\dagger$] Correspondence to: apn2108@columbia.edu, darhodes@wisc.edu
\end{affiliations}

\begin{abstract}
%
Achieving electrostatic control of quantum phases is at the frontier of condensed matter research. Recent investigations have revealed  superconductivity tunable by electrostatic doping in twisted graphene heterostructures and in two-dimensional (2D) semimetals such as WTe$_2$ \cite{twisted_bilayer, twisted_double_bilayer, twisted_trilayer,wte2_superconductor_1, wte2_superconductor_2}. Some of these systems have a polar crystal structure that gives rise to ferroelectricity, in which the interlayer polarization exhibits bistability driven by external electric fields \cite{graphene_ferro, wte2_ferro_2, wte2_ferro}. Here we show that bilayer $T_d$-MoTe$_2$ simultaneously exhibits ferroelectric switching and superconductivity. Remarkably, a field-driven, first-order superconductor-to-normal transition is observed at its ferroelectric transition. Bilayer $T_d$-MoTe$_2$ also has a maximum in its superconducting transition temperature ($T_\textrm{c}$) as a function of carrier density and temperature, allowing independent control of the superconducting state as a function of both doping and polarization. We find that the maximum $T_\textrm{c}$ is concomitant with compensated electron and hole carrier densities and vanishes when one of the Fermi pockets disappears with doping. We argue that this unusual polarization-sensitive 2D superconductor is driven by an interband pairing interaction associated with nearly nested electron and hole Fermi pockets.
\end{abstract}
%
\indent Ferroelectricity has been recently found in a number of two-dimensional (2D) van der Waals layered heterostructures that break inversion symmetry either intrinsically \cite{wte2_ferro,wte2_ferro_2} or through heterostructure engineering \cite{graphene_ferro}. In contrast to traditional ferroelectricity that arises due to long range Coulomb interactions in compounds such as BaTiO$_3$\cite{rabe2007modern}, this phenomenon is thought to emerge due to the interplay between interlayer sliding and the small dipole moments arising from broken inversion symmetry. For instance, in bilayer hBN, out-of-plane electric fields can cause interlayer sliding, changing the stacking order from BA to AB and switching the polarization direction\cite{bn_ferro}. The same principle has been extended to rhombohedral-stacked bilayer transition metal dichalcogenides (TMDs) \cite{wse2_ferro, wse2_ferro_2}, and orthorhombic-stacked bilayer $T_d$-WTe$_2$\cite{wte2_ferro,wte2_ferro_2}, demonstrating a viable path to achieving ferroelectric behavior in almost any noncentrosymmetric 2D heterostructure. Compared to thin film oxides (\textit{e.g.}, BiFeO$_3$)\cite{sando2014bifeo3}, ferroelectric 2D heterostructures offer compelling advantages: tunable electronic behavior via conventional electrostatic techniques, modifications through strain, and the ability to exploit ferroelectricity to control other electronic states. Importantly, because 2D ferroelectric structures are atomically thin, in-plane metallic states are compatible with the out-of-plane polarization\cite{wte2_ferro_2}. For example, several metallic TMDs show 2D superconductivity at low temperatures\cite{ye2010liquid,hamill2021two,mote2_supercondcutor_mono}. Ferroelectricity thus offers another tuning knob, besides electrostatic doping, to control and assess 2D superconductivity. A possible candidate to achieve this goal is few-layer $T_d$-MoTe$_2$ (hereafter referred to as MoTe$_2$), which has been independently shown to display ferroelectricity and compensated superconductivity \cite{mote2_ferro_pfm,mote2_supercondcutor_mono}.\\
\indent In the bulk, MoTe$_2$ is a nearly compensated semimetal\cite{mote2_weyl, mote2_weyl_2} with a superconducting \tc of 100 mK \cite{qi2016superconductivity, wang2020evidence}. Density functional theory (DFT) calculations suggest that monolayer\cite{mote2_supercondcutor_mono} and bilayer MoTe$_2$ (see Fig. 1b) retain this charge compensated behavior, displaying nearly compensated electron and hole pockets at the Fermi level with a small bilayer splitting. The superconducting transition temperature unusually increases with decreasing thickness, reaching a maximum of $\sim$7 K in the monolayer limit\cite{mote2_supercondcutor_mono}. This behavior is distinctly different from other 2D superconductors, where superconductivity is dominated by a single carrier type, such as monolayer WTe$_2$\cite{wte2_superconductor_1,wte2_superconductor_2} or few-layer NbSe$_2$\cite{xi2016ising}. MoTe$_2$ also takes on a polar crystal structure, where net out-of-plane polarization arises between layers in the few-layer limit. This polarization, along with its out-of-plane switching, has been previously demonstrated \textit{via} piezoresponse force microscopy measurements \cite{mote2_ferro_pfm}. The switching behavior, similar to that seen in WTe$_2$, has been attributed to interlayer sliding\cite{wte2_ferro_reason,wte2_ferro_reason2}. Assuming this to be the case, bilayer MoTe$_2$ is the thinnest possible material that still has the sliding degree of freedom present, and we use it as a platform to study the interaction between the superconducting state and the electric polarization.\\
\indent For our experiments, we fabricate bilayer MoTe$_2$ samples with dual top and bottom gate electrodes whose voltages $V_T$ and $V_B$ allow us to tune carrier-density, $\Delta n = \epsilon_{hBN}\varepsilon_o(V_T/d_T + V_B/d_B)/e$, and displacement field, $D = \epsilon_{hBN}(V_T/d_T - V_B/d_B)/2$, independently (Fig. 1a and Supplementary Fig. 2). Shown in Fig. 1c are measurements of the Hall resistance at 250 mK for several different carrier densities. The data at low $\Delta n$ clearly shows that the Hall effect is nonlinear in field. As we increase $\Delta n$, we see the Hall signal evolve from a nonlinear to linear, suggesting a single dominant carrier at high $\Delta n$ (a full discussion of the Hall effect in our device is provided in Supplementary Information (SI)). We also see a non-saturating magnetoresistance at zero doping in our samples, also shown in Fig. 1d. All of these features are hallmarks of compensated semimetals\cite{MR_semimetals, mote2_magnetoresistance, wte2_magnetoresistance}. These measurements are, therefore, broadly consistent with the expected electronic structure of pristine material from density functional theory (DFT) calculations. This agreement indicates that there is no large external doping or degradation present in our bilayer samples.\\ 
\indent Figure 1e shows the resistance of the undoped sample as a function of temperature. A clear superconducting transition is observed, with a \tc of 2.3 K \cite{mote2_supercondcutor_mono}. In order to probe the presence of ferroelectricity, we sweep $D$ at a fixed density in the normal state (4 K). We observe hysteretic switching of the four-probe resistance, R$_{xx}$, as shown in Fig. 1f. While such resistive switching has not previously been observed in MoTe$_2$, it has been seen in multilayered WTe$_2$ \cite{wte2_ferro, wte2_ferro_2}. In analogy with WTe$_2$, we associate this resistance bistability with an interlayer sliding transition, which flips the out-of-plane polarization\cite{wte2_ferro_reason, wte2_ferro_reason2}.\\
\indent To study how the ferroelectricity interacts with 2D superconductivity at low temperatures, we measure $R_{xx}$ as a function of displacement field at 1.7 K, as shown in Fig. 2a. We clearly observe the presence of both superconductivity and hysteretic switching due to ferroelectricity. Starting from $D<-2$V/nm (blue curve, Fig. 2a), both the displacement field and the sample polarization point in the same direction. On decreasing the magnitude of the displacement field and then flipping its sign, superconductivity emerges gradually - resulting in a drop of the sample resistance to zero. The sample remains superconducting until the displacement field switches the crystal's internal polarization, at which point it transitions to the normal state. The behavior on the downward sweep of displacement field is similar, with the hysteresis expected from the ferroelectricity. The asymmetry of the superconducting butterfly loop with applied field has been seen in other van der Waals ferroelectrics\cite{wte2_ferro,wte2_ferro_2,bn_ferro}. Its origin is likely extrinsic to the material itself and is possibly related to substrate and contact asymmetries.\\ 
\indent The coexistence of ferroelectric switching and superconductivity in a single material can be used to make a superconducting switch driven by external electric field. We illustrate this in Fig 2b, which shows the resistance of the sample as a function of time as external fields are applied to the sample. Starting from the normal (superconducting) state, an electric field pulse of appropriate positive (negative) sign can drive a transition  to the superconducting (normal) phase. Once the switching between states is established, the resistance of the sample continues to stay in the new state indefinitely, as is seen in the figure. Such a first-order switch for superconductivity may lead to low-temperature classical and quantum electronics applications, such as low-power transistors and tunable qubits in the future.\\ 
\indent In order to explore the connection between ferroelectric switching and superconductivity more carefully, we perform displacement field sweeps at different temperatures, the results of which are summarized in Fig. 2c for the two sweep directions. For the forward direction, as the displacement field is lowered from a high absolute value to a low value, a superconducting transition emerges at low temperatures. On continuing to sweep the displacement field through zero, \tc continues to increase until the displacement field switches the polarization, at which point superconductivity is lost. In both sweep directions, the maximum \tc is therefore seen just before a switching event. This continuous tuning of \tc prior to switching shows that the mechanism for the superconducting state is intimately tied to the internal electric field of the sample.\\
\indent The ferroelectric switching behavior described above is density dependent, as is the observed superconducting behavior. In order to understand the region of ferroelectricity in the carrier doping versus displacement field phase diagram, we perform sweeps of the displacement field similar to Fig. 2a at various doping levels, both in the superconducting state and in the normal state. Taking the difference in resistance between the forward and backward sweep directions identifies when hysteretic switching is observed in the samples. Shown in Fig. 2d are the results of such measurements in the normal state ($T$ = 8 K) and in the superconducting state ($T$ = 1.5 K). Focusing on the normal state, we observe that the hysteretic switching in our samples is limited to a doping range of $\Delta n = \pm 2\times10^{13}$ cm$^{-2}$ in our measurements. Such switching behavior is observed up to 60 K (see SI). Whether the absence of ferroelectric switching at high doping is an intrinsic effect or whether it is due to constraints of the gate voltages we can achieve in our experiment is currently unclear. \\
\indent The corresponding measurements in the superconducting state clearly show the role of the internal electric field in the observed superconducting behavior. In general, we observe that electric field can drive a superconducting transition in the sample both at low and high doping (black dashed lines, Fig. 2e). We see that  hysteretic superconductivity occurs in parts of the phase diagram that are close to the boundary where field-driven superconductivity intersects with the normal state ferroelectric behavior. This phenomenology suggests the simple hypothesis that it is the total internal electric field in the sample that controls superconductivity: the sample polarization, when flipped, can turn the superconducting phase on or off if sufficiently close to the field-driven superconducting transition.\\
\indent So far, we have discussed the displacement field dependent properties of the sample. Now, we turn to the carrier density dependence of superconductivity in the sample at zero displacement field. Shown in Fig. 3a is a color plot of the temperature-dependent resistance of the sample that shows the presence of a maximum in $T_\textrm{c}$ of 2.5 K. By fitting the Hall measurements in Fig. 1c with a two-band semiclassical model (see SI for details), we extract the independent electron and hole carrier densities of the bilayer (Fig. 3b). This plot shows that the maximum $T_\textrm{c}$ is closely correlated to the compensation point of the material, $\Delta n = 0$. To further confirm our Hall analysis and the compensated behavior, we measure the magnetoresistance as a function of doping. In agreement with the two-band model, we see a non-saturating magnetoresistance for $\Delta n = 0$ and for intermediate values of $\Delta n$  where both electron and holes bands cross the Fermi level (see SI).\\
Upon doping with electrons we see an overall reduction in $T_\textrm{c}$ until $\Delta n = 2\times10^{13}/\textrm{cm}^2$ is reached, beyond which superconductivity disappears completely (down to 250 mK). This complete suppression of superconductivity occurs at the same concentration where the Hall effect becomes linear and the magnetoresistance saturates with increasing magnetic field. Therefore, superconductivity disappears when the chemical potential is raised above the hole pocket. While the same trend is seen upon doping with holes, we cannot suppress superconductivity completely since we are limited in the voltage we can safely apply to our gates. The main finding that \tc is maximized near the compensation point is reproduced in multiple devices (see Extended Data Fig. 1), but the precise shape of the \tc versus doping curve shows device to device variations.\\
\indent At zero displacement field, it is clear from the discussion above that both hole and electron carriers are required to maximize superconductivity. We can ask whether the same is true when hysteretic superconducting switching is observed. To study this, we measure the Hall effect in the region of ferroelectric hysteresis at $\Delta n = 1.5\times10^{13}/\textrm{cm}^2$. Shown in Fig. 3c are a subset of this data as the displacement field is swept from right to left (left panel) and from left to right (right panel). The curves are color-coded for different displacement field values according to the symbols in Fig. 3d. Starting from high positive displacement field, where the sample is in the normal state, the Hall resistance is linear in field and is well-fitted using a single electron band model. As we sweep the field down and enter the superconducting state, the Hall effect develops a pronounced nonlinearity and shows the presence of both types of carriers. Upon switching out of the superconducting state at large negative field, the Hall effect goes back to nearly linear, showing that the hole carrier density is sharply reduced. The same trend is seen when sweeping the field back from negative to positive values, with a non-linear Hall effect seen where the sample is superconducting. The extracted carrier concentrations are shown in Fig. 3d, with the hysteretic superconducting regions shaded in red (for the sweep from right to left) and in blue (from left to right). The observation that we need both types of carriers to have superconductivity is clearly seen from this figure.\\
\indent To further investigate the metallic state from which 2D superconductivity emerges, we measure the temperature dependence of the normal state resistance at various doping levels and zero displacement field, as shown in Extended Data Fig. 2 a,b. In bulk MoTe$_2$, the temperature dependence of the resistance is $T^2$ up to 50 K, after which it becomes dominated by phonon scattering\cite{zandt2007quadratic}. On the other hand, for bilayer MoTe$_2$, we observe an approximately linear-in-$T$ resistance for intermediate temperatures. For most doping values, the resistance recovers the usual $T^2$ dependence at low temperatures, as shown in Extended Data Fig. 2c. However, as summarized in extended data Fig. 2c, for a narrow range of $\Delta n$ near the edge of the superconducting dome, the $T$-linear behavior persists down to the superconducting $T_\textrm{c}$. This behavior has been qualitatively reproduced in at least one other device (see Extended Data Fig. 3). In certain unconventional superconductors\cite{hussey2018tale,greene2020strange} and other 2D correlated systems\cite{cao2020strange,ghiotto2021quantum}, a $T$-linear resistance is often associated with an underlying strange metallic state, whose origin remains widely debated. In the present case of bilayer MoTe$_2$, it is important to notice that the $T$-linear contribution is comparable to the residual resistance.\\ 
\indent The fact that superconductivity appears only when both electron and hole pockets are present suggests that interband processes are behind the pairing mechanism. The situation is reminiscent of iron-pnictide superconductors, which are also compensated semimetals \cite{fernandes2022iron}. In that case, the pairing interaction has been proposed to be enhanced by spin fluctuations. Interestingly, prior results on monolayer MoTe$_2$ hypothesized that the enhancement in $T_\textrm{c}$ was unconventional and likely related to enhanced spin fluctuations\cite{mote2_supercondcutor_mono}. To further analyze this scenario, we compute the leading eigenvalue $\lambda_{\mathrm{max}}$ of the multi-orbital Lindhard susceptibility from DFT calculations (see Methods for details). To set the stage, we first consider the monolayer case, whose Fermi surface is shown in Fig. 4a. As seen in Figs. 4b-c, $\lambda_{\mathrm{max}}$ is peaked at the two nesting wave-vectors $q_{1M}$ and $q_{2M}$ that connect the edges of the electron and hole pockets along the $k_x$ axis.\\
\indent The situation in bilayer MoTe$_2$ is slightly more complicated due to the presence of an additional electron pocket along the $k_y$ direction as well as the bilayer splitting (Fig. 4d). As a result, $\lambda_{\mathrm{max}}$ displays a peak not only at the nesting wave-vector, $q_\textrm{1B}$, connecting the hole and electron pockets approximately along $k_x$, but also at the wave-vector, $q_\textrm{2B}$, connecting the hole and electron pockets along $k_y$ (Fig. 4e). Moreover, an additional peak appears at $q_\textrm{3B}$, which connects the bilayer-split hole pockets. Electron-electron interactions involving the same or different orbitals are expected to enhance one or more of these peaks, thus providing a possible mechanism for the interband pairing interaction. This would imply that MoTe$_2$ is a rather unusual 2D superconductor. While other mechanisms cannot be ruled out at present, this simple nesting-driven scenario is qualitatively consistent with the doping dependence of \tc observed in bilayer MoTe$_2$. Indeed, as shown in Fig. 4f, the peak of $\lambda_{\mathrm{max}}$ at $q_\textrm{1B}$ first increases for small electron doping and then decreases and disappears once the hole pocket is pushed down below the Fermi level. \\
\indent Our finding of a tunable \tc in the ferroelectric regime of bilayer MoTe$_2$ shows that this material is a promising platform for controlling an unusual type of 2D superconductivity with two independent and highly-precise knobs: doping and displacement field. At the same time, the discrete switching of superconductivity at the ferroelectric transition opens up new possibilities for quantum devices with a first order switch for the superconducting phase. We expect that these effects will have interesting dependence on layer thickness\cite{wte2_ferro} and twist angle that tunes the degree of inversion symmetry breaking\cite{bn_ferro}. Such effects should also be present in other noncentrosymmetric 2D superconductors\cite{zhai2022prediction}. Finally, the control of these properties using ultrafast electromagnetic excitations that couple to the lattice is also an appealing prospect \cite{li2019terahertz,ji2021manipulation}.

\clearpage


\begin{figure*}[t] 
    \centering
	\includegraphics[width=1.0\linewidth]{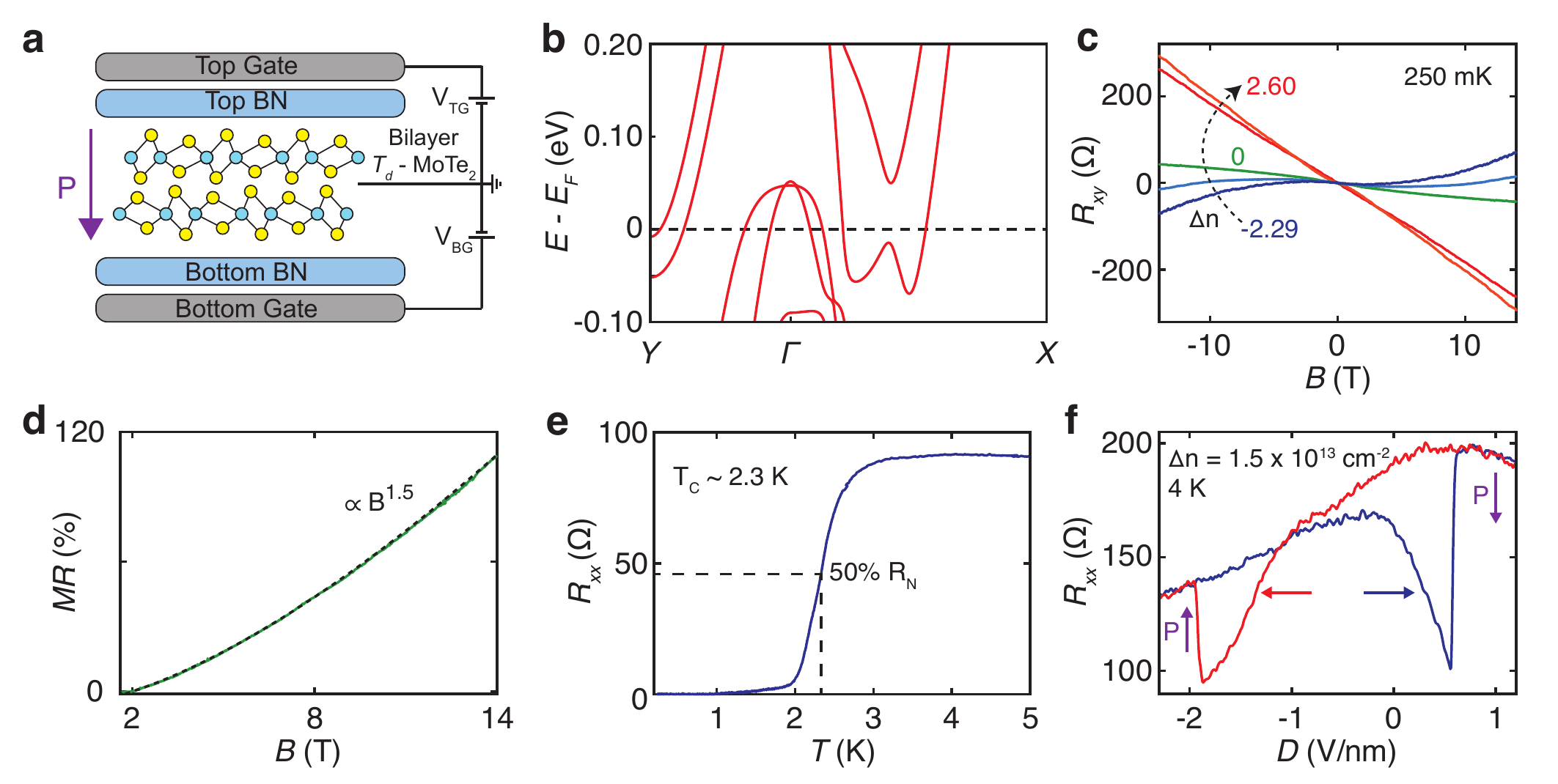}
        \caption*{\textbf{Fig. 1 $|$ Electronic properties of 2L $T_d$-MoTe$_2$. a,} Dual-gated device schematic. $\Vec{P}$ denotes the direction of internal polarization in the crystal. \textbf{b,} DFT calculated electronic band structure for bilayer MoTe$_2$. \textbf{c,} Evolution of Hall resistance, $R_{xy}$, with electrostatic doping. $\Delta$n is in units of 10$^{13}$ cm$^{-2}$. \textbf{d,} Intrinsic ($\Delta n = 0$) non-saturating magnetoresistance in bilayer MoTe$_2$; a sub-quadratic magnetic-field dependence is fit to data (black). \textbf{e,} Superconducting transition in bilayer MoTe$_2$ for $\Delta n = 0$. \textbf{f,} Ferroelectric switching in bilayer MoTe$_2$ with an applied displacement field.}
\label{Fig. 1}
\end{figure*}

\begin{figure*}[t] 
	\centering
        \includegraphics[width=0.7\textheight, keepaspectratio]{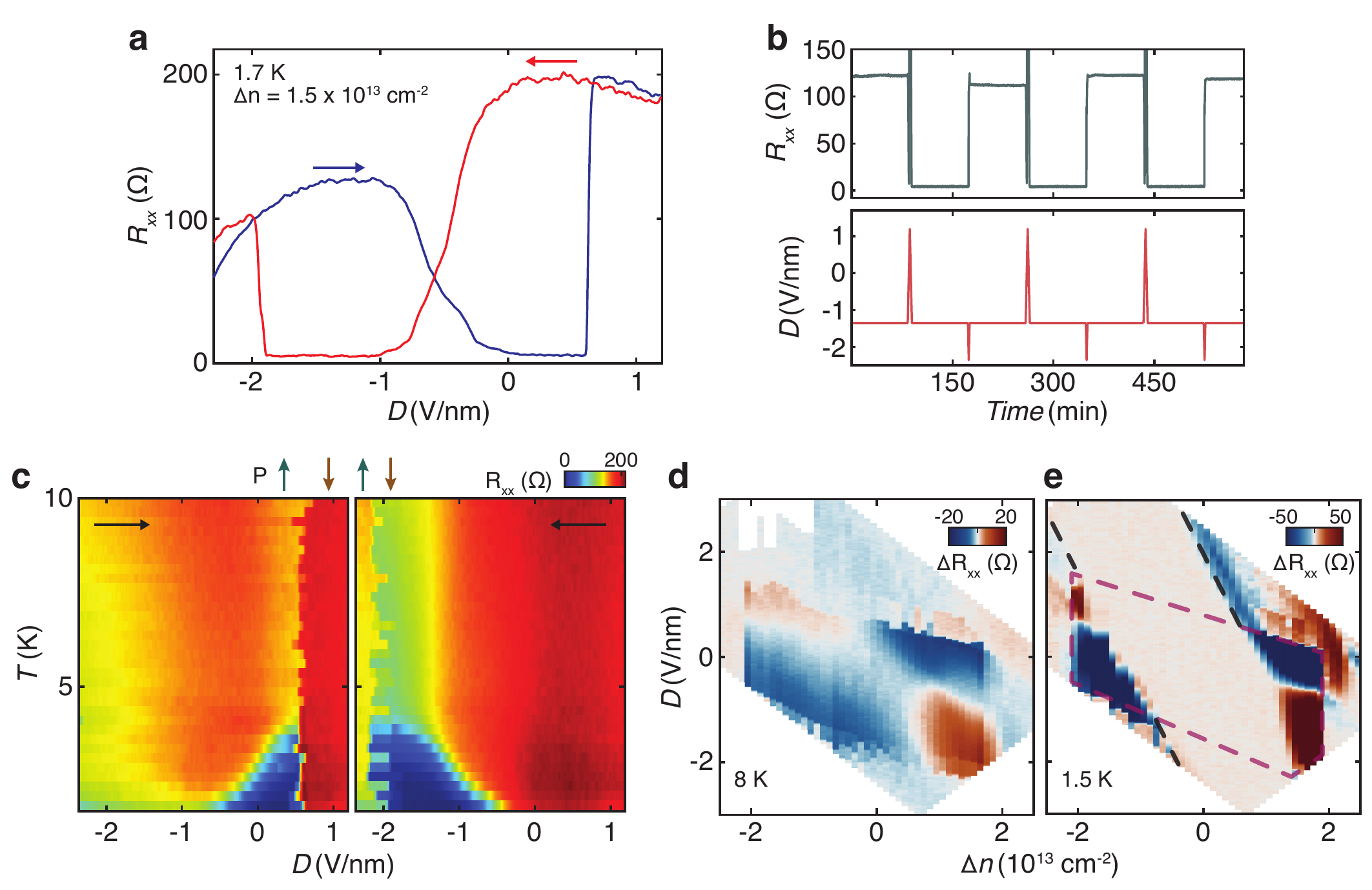}
	\caption*{\textbf{Fig. 2 $|$ Coupled ferroelectricity and superconductivity in 2L $T_d$ - MoTe$_2$. a,} Butterfly loops with bistable normal and superconducting states, indicating coupled ferroelectric and superconducting states. Red and blue arrows denote the direction of $D$ sweep. \textbf{b,} Reversible switching between superconducting and normal states at a fixed charge carrier density as shown in (a). \textbf{c,} Temperature evolution of the coupled ferroelectric and superconducting behavior at $\Delta n = 1.5 \times$10$^{13}$ cm$^{-2}$. A dramatic increase in $T_\textrm{c}$ is observed just before ferroelectric switching. The internal polarization of the crystal is marked with arrows on top. \textbf{d,} Resistance difference between displacement field sweep directions shows regions of ferroelectricity at 8 K and its evolution with doping. \textbf{e,} At 1.5 K, highlighting the regions of superconductivity. The largest difference in resistance occurs at dopings with hysteretic superconductivity. The magenta dashed line represents the hysteretic region in the normal state. The black dashed line represents the boundary of the field-driven superconductivity.}
\label{Fig. 2}
\end{figure*}

\begin{figure*}[t] 
    \centering
	\includegraphics[width=1.0\linewidth]{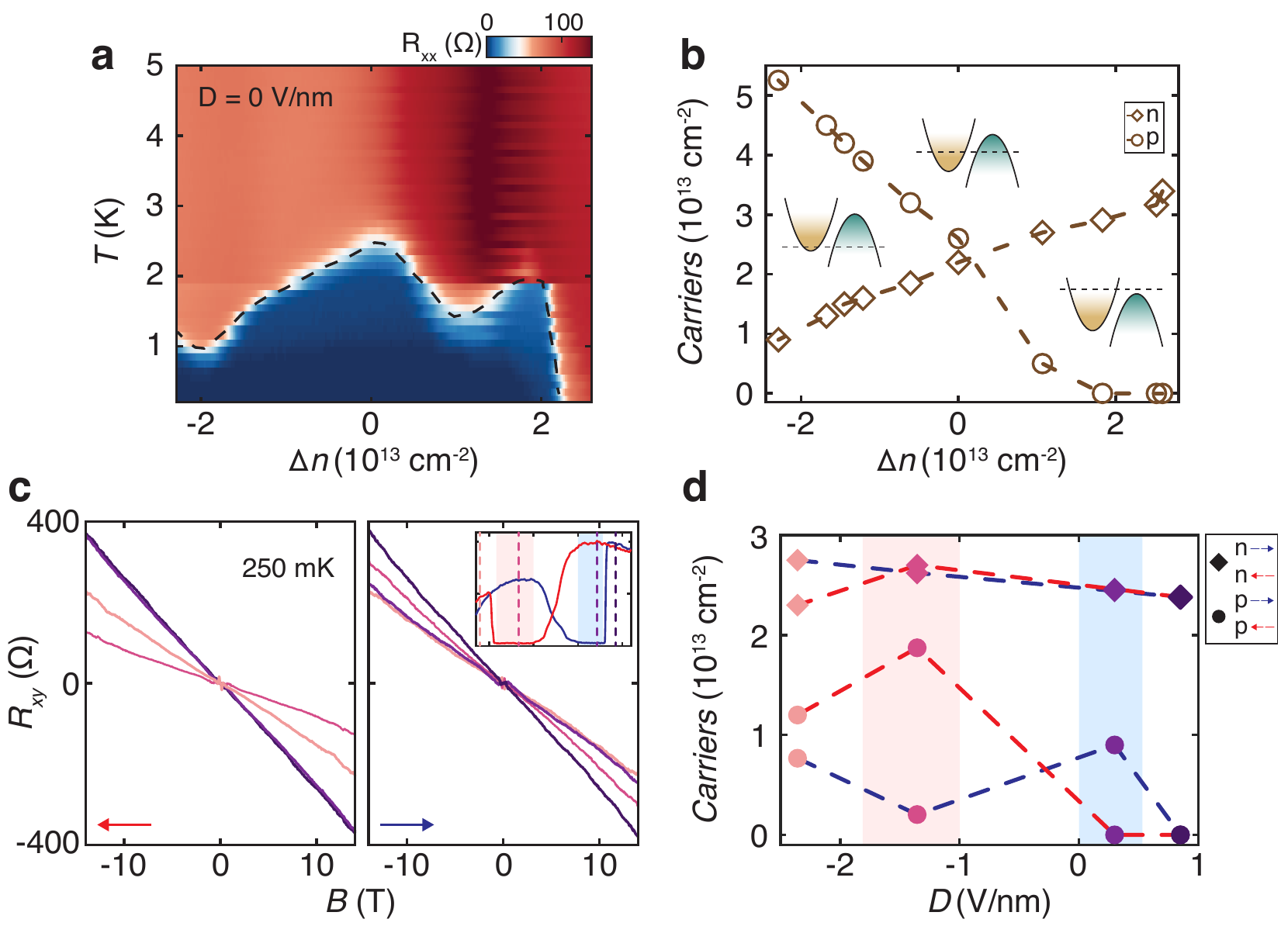}
	\caption*{\textbf{Fig. 3 $|$ Doping-dependent superconducting properties of 2L $T_d$ - MoTe$_2$. a,} Evolution of a superconducting dome with electrostatic doping and temperature. \textbf{b,} Carrier concentration as a function of $\Delta n$, as extracted by fitting a two-band model to the Hall resistance. \textbf{c,} Evolution of Hall resistance at a fixed doping, $\Delta n = 1.5 \times$10$^{13}$ cm$^{-2}$, while varying $D$ (inset)). \textbf{d,} Extracted carrier concentrations from the two-band model to Hall data in (c). The red and blue dashed lines denote  sweeping $D$ right to left, and left to right, respectively. Diamond and circle colors correspond to their respective Hall response as plotted in (c).}
\label{Fig. 3}
\end{figure*}

\begin{figure*}[t] 
	\includegraphics[width=1.0\linewidth]{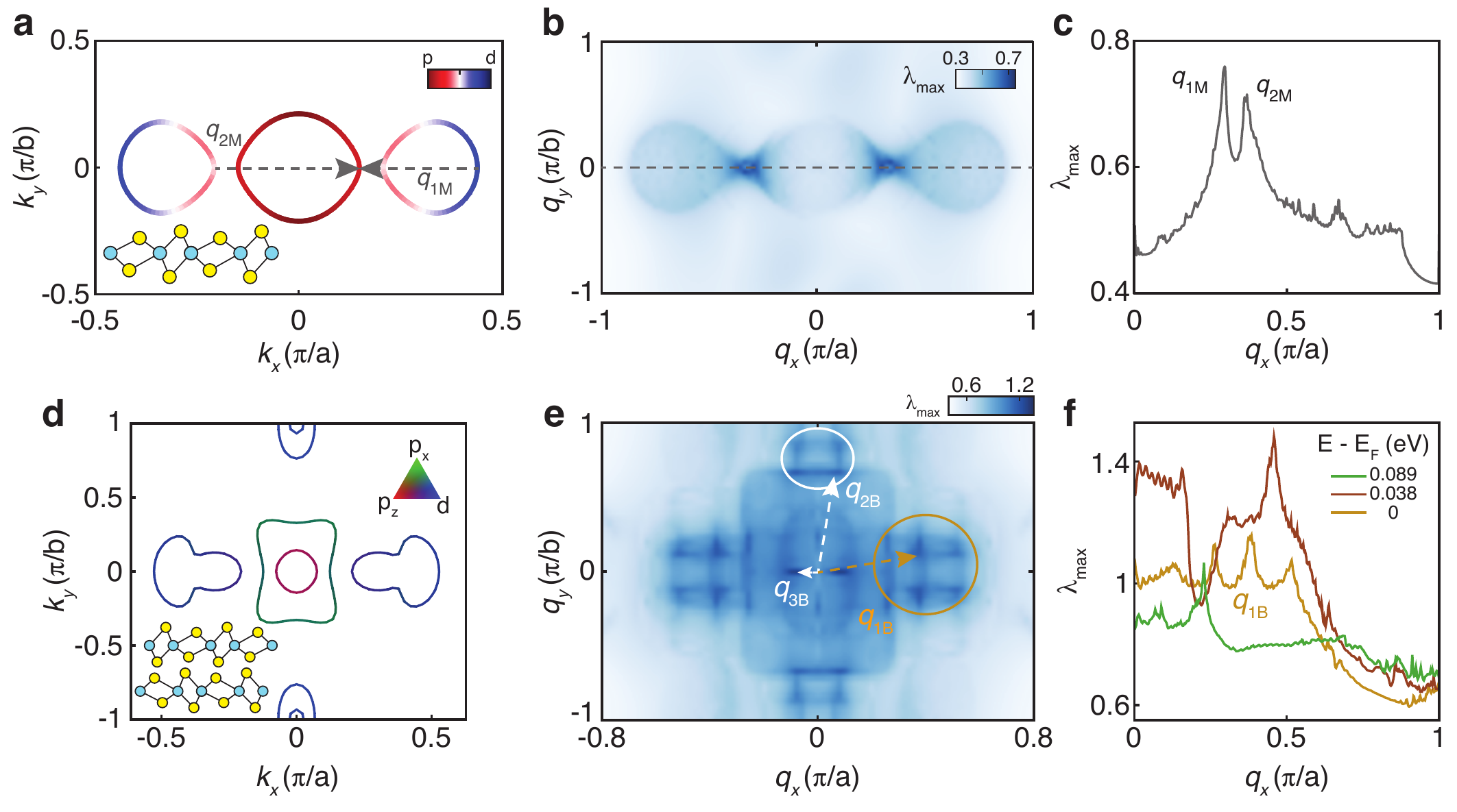}
	\caption*{\textbf{Fig. 4 $|$ Fermi surface nesting and superconductivity in MoTe$_2$ a,} Fermi surface with nesting vectors for monolayer MoTe$_2$. Colors of the Fermi surface correspond to orbital projections. \textbf{b,} Lindhard susceptibility values, $\lambda_\textrm{max}$ for monolayer MoTe$_2$. \textbf{c,} Linecut of $\lambda_\textrm{max}$ from (b) along the $q_x$ direction with $q_y$ = 0. Peaks correspond to interband nesting vectors for electron-hole pockets on the Fermi surface. \textbf{d,} DFT calculated Fermi surface for bilayer MoTe$_2$ showing hole pocket centered around $\Gamma$ with electron pockets along the $\Gamma$-X and $\Gamma$-Y directions. Orbital projections have been depicted in color with the hole-pocket being $p_x$ and $p_z$ while electron pocket as $d_{x^2-y^2}$. \textbf{e,} $\lambda_\textrm{max}$ values for bilayer MoTe$_2$. Local maxima in $\lambda_\textrm{max}$ correspond to nesting vectors. $\Vec{q}_\textrm{1B}$ corresponds to the electron-hole nesting vector. \textbf{f,} $\lambda_\textrm{max}$ as a function of doping. Colours correspond to doping along the band structure as mentioned in legend.}
\label{Fig. 4}
\end{figure*}

\clearpage

\begin{methods}
\subsection{Crystal Growth}
Single crystals were grown by a self-flux method using excess Te. Mo powder, 99.9975\%, was loaded into a Canfield crucible with Te, 99.9999\% lump in a ratio of 1:25, respectively, and subsequently sealed in a quartz ampoule under vacuum ($\sim$ 1 mTorr). Afterwards, the ampoule was heated to 1120 $^\circ$C over 12 hrs, and then held at this temperature for 5 days before cooling down to 880 $^\circ$C over a period of 3 weeks and subsequently centrifuging to remove excess Te. After removing crystals from the original quartz ampoule, crystals were again resealed under vacuum in a quartz ampoule and annealed in a temperature gradient with the crystals held at 435 $^\circ$C and the cold of the ampoule held at room temperature for 2 days. The final annealing is used to rid the crystals of any interstitial Te and is critical for obtaining the highest residual resistivity ratios.

\subsection{Device Fabrication}
Few-layered MoTe$_2$ is extremely air-sensitive and degrades within a matter of minutes in the ambient \cite{gan2020bandgap}. Therefore, bilayer $T_d$-MoTe$_2$ devices were fabricated in a nitrogen-filled glovebox with H$_2$O and O$_2$ levels below 0.5 ppm and electrical contact made utilizing pre-patterned electrodes fully encapsulated in hBN (top hBN = 29 nm, bottom hBN = 7 nm) on a metal back gate. First, local backgates were defined on Si/SiO$_2$ substrates using traditional e-beam lithography techniques (EBL), metal deposited using e-beam deposition (2/10 nm Ti/Pd), and then cleaned via ultrahigh vacuum annealing at 300 $^\circ$C and subsequent exposure to a low power O$_2$ plasma for 10 minutes. Thin hBN (7-10 nm) was dry transferred using either PPC or PC onto the metal backgates. This was followed with another EBL step to define pre-patterned contacts that did not extend beyond the hBN edge and e-beam deposition (2/12 nm Ti/AuPd). Pre-patterned contacts were then cleaned using an atomic force microscope (AFM) tip in contact mode (force $\sim$300 nN, tip radius 7 nm) and loaded into a nitrogen or argon-filled glovebox. Few-layer MoTe$_2$ was mechanically exfoliated onto a polydimethylsiloxane stamp and then transferred onto Si/SiO$_{2}$ substrates. Bilayers were then identified via optical contrast on the Si/SiO$_{2}$ substrates, and later confirmed by measuring $R_\textrm{xx}$, and, where possible, step height in AFM. To complete the heterostructure, hBN (20-30 nm) was then picked up via a dry stacking method\cite{wang2013one} using polypropylene carbonate (PPC) as the polymer. The picked-up hBN was then used to pickup bilayer MoTe$_2$ and finally transferred on to the bottom hBN/pre-patterns at 120 $^\circ$C. Afterwards, through-holes were defined in a polymethyl methacrylate (PMMA) resist (950k A6) and etched using EBL and reactive ion etching with SF$_6$/O$_2$ plasma. A final EBL and e-beam deposition step was used to define and deposit a top gate and contacts to the pre-patterns (2/30/90 nm, Ti/Pd/Au).   
\subsection{Electrical Transport}
Electrical transport measurements of our devices were performed either in a 3He cryostat equipped with a superconducting magnet (14 T) or in a 4He variable temperature insert with base temperature of 1.7 K. Standard lock-in measurements were taken with an AC excitation (37.77 Hz) of 10-100 nA using an SR860 or SR830 lock-in amplifier with a 1 M$\Omega$ resistor in series ($V_\textrm{bias} = 10 - 100 mV$). For measurements pertaining to superconductivity, passive RC filters with a cutoff frequency of $\sim$50 kHz that were kept at $\sim$4 K near the sample space were utilized on both current and voltage contacts.
\subsection{Theoretical Calculations}
First-principles DFT calculations were performed with the projector augmented wave formalism as implemented in the Vienna \textit{ab initio} simulation package \cite{Kresse199607, Kresse199610,Blochl1994,Kresse1999}.
PBEsol exchange correlation functional \cite{Perdew2009} was used with a 360 eV cutoff energy for the plane-wave basis.
The in-plane lattice parameters for monolayer and bilayer MoTe$_2$ structures were fixed to $a = 6.33$ {\AA} and $b = 3.469$ {\AA}, obtained experimentally from bulk structure \cite{Brown1966}, and the internal ionic coordinates were relaxed. 
A shifted Monkhorst-Pack grid \cite{Monkhorst1976} of $9\times18\times3$ $k$-points was used for the monolayer and $8\times16\times2$ $k$-points was used for the bilayer primitive cell. 
Spin-orbit coupling was not included since we found that it did not generate significant qualitative differences in the Fermi surface.
The WANNIER90 package \cite{Mostofi2008,Marzari2012} was used to calculate maximally localized Wannier functions (MLWF). From those, a 4-band tight-binding model was obtained for the monolayer, with 2736 in-plane nearest-neighbor hopping parameters involving the $p_x$ ($d$) orbitals of the two inequivalent Te (Mo) atoms in the unit cell. Note that not all hopping parameters are independent due to symmetry constraints. A 10-band tight-binding model was constructed for the bilayer, with two $p_z$ orbitals added to account for the inter-layer interaction, resulting in 15300 in-plane nearest-neighbor hopping parameters. The static multi-orbital Lindhard susceptibility tensor $\chi^{\alpha \beta \gamma \delta}(\mathbf{q})$ was defined according to \cite{Graser2009}:
\begin{equation}
\chi^{\alpha\beta}_{\gamma\delta}(\mathbf{q}) = -\frac{1}{N}\sum_{\mathbf{k},m n}\frac{a^\gamma_{m}(\mathbf{k})a^{\alpha*}_{m}(\mathbf{k})a^\beta_{n}(\mathbf{k+q})a^{\delta*}_{n}(\mathbf{k+q})}{E_{n}(\mathbf{k+q})-E_{m}(\mathbf{k})+i0^+}\left[f(E_{n}(\mathbf{k+q}))-f(E_{m}(\mathbf{k}))\right]
\end{equation}where $a_{m}^{\gamma}(\mathbf{k})$ are eigenvector components associated with the change from orbital basis (Greek letters) to band basis (Latin letters),  $E_m(\mathbf{k})$ are the energy eigenvalues in band basis, and $f(E)$ is the Fermi function, which at  $T=0$ becomes the theta function $\theta(E-E_F)$. The eigenvalue problem \cite{Christensen2016} $\chi^{\alpha \beta \gamma \delta} v_{\alpha \beta}^{(n)} = \lambda^{(n)} \, v_{\gamma \delta}^{(n)}$ was then solved to find the maximum eigenvalue $\lambda_{\mathrm{max}} \equiv \max{(\lambda^{(n)})}$ for each momentum value with PythTB \cite{PythTB} and in-house scripts, using a $50\times100$ grid for the internal momentum sum.

\end{methods}

\clearpage

\begin{addendum}
\normalsize
 \item  \hspace{- 2mm} We thank Andrew Millis for discussions. The experimental portion of this research was primarily supported by the NSF MRSEC program through Columbia University in the Center for Precision-Assembled Quantum Materials under award number DMR-2011738 (fabrication, measurements, and data analysis). A.S., T.B., and R.M.F. (theoretical modeling) were supported by the National Science Foundation through the University of Minnesota MRSEC (DMR-2011401). D.A.R. and Z.L. (growth, measurements, and data analysis) were supported by the University of Wisconsin-Madison, Office of the Vice Chancellor for Research and Graduate Education with funding from the Wisconsin Alumni Research Foundation. D.A.R. was partially supported by the NSF MRSEC program through the University of Wisconsin-Madison under award number DMR-1720415. A.N.P. acknowledges salary support from the NSF via grant DMR-2004691, from AFOSR via grant FA9550-21-1-0378 by the ARO-MURI program with award no. W911NF-21-1-0327. K.W. and T.T. acknowledge support from the Element Strategy Initiative conducted by the MEXT, Japan (grant no. JPMXP0112101001), and JSPS KAKENHI (grant nos. 19H05790, 20H00354 and 21H05233). 
 
\item[Author contributions] The experiment was designed by A.J., D.A.R., and A.N.P. Devices were fabricated by A.J., D.A.R., and Z.L. A.J., C.R.D., and D.A.R. performed the measurements, and analyzed the data. A.S. developed theoretical models and performed calculations supervised by T.B. and R.F.M. T.T. and K.W. supplied hBN single crystals, and D.A.R. and J.C.H. synthesized MoTe$_2$ single crystals. D.A.R., A.J., and A.N.P. wrote the manuscript with the input of all other authors.    

 \item[Corresponding Authors] Correspondence and requests for materials
should be addressed to A.N.P (email: apn2108@columbia.edu), or D.A.R. (email: darhodes@wisc.edu).

 \item[Ethics Declaration] The authors declare no competing interests.
 
 \item[Data Availability] Datasets used to construct plots and support other findings in this article are available from the corresponding author upon request.
\end{addendum} 
\newpage

\section*{Extended data figures and tables}

\begin{figure*}[ht]
	\includegraphics[width=1.0\linewidth]{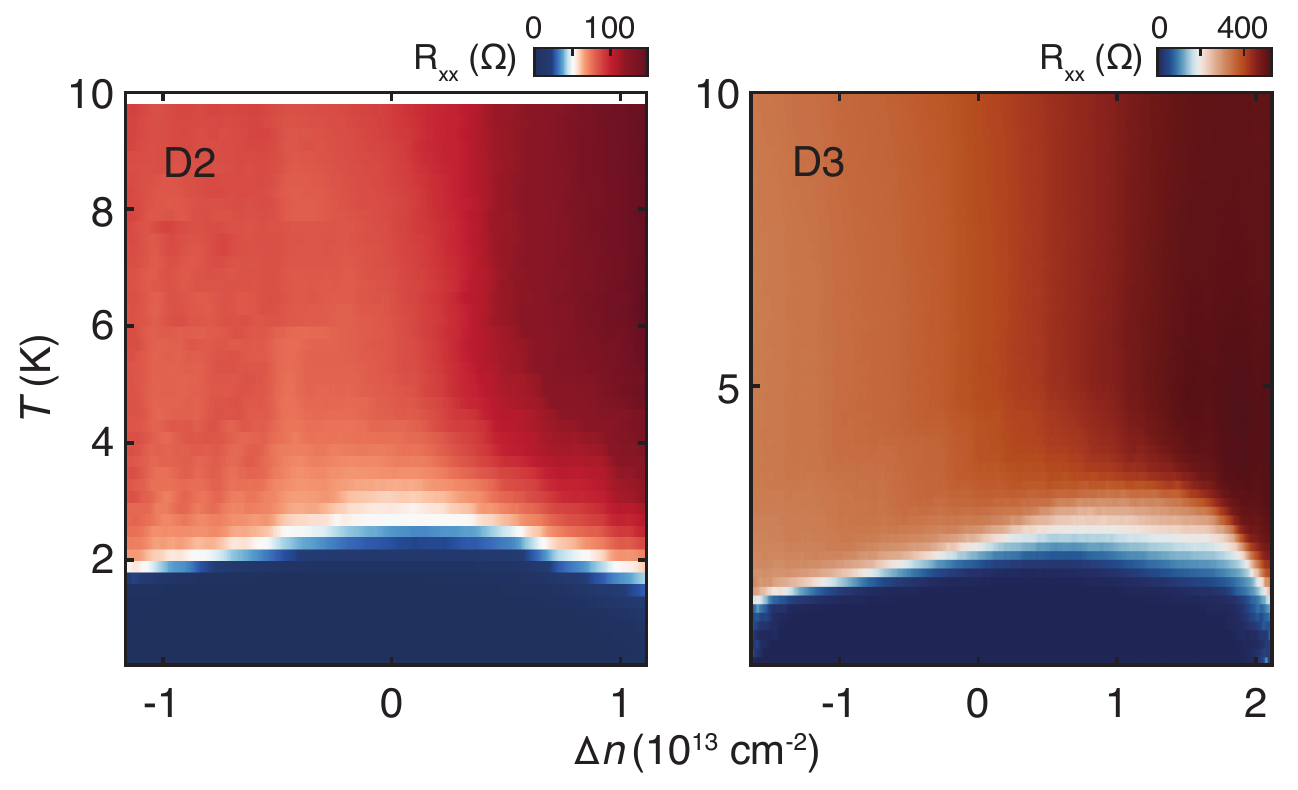}
	\caption*{\textbf{Extended Data Figure 1 $|$}Superconducting dome observed in two other devices at D = 0 V/nm. Limitations due to the dielectric strength of the $h$BN in these devices precluded us from reaching higher doping and, consequently, the formation of a complete superconducting dome.}
\label{Ext1}
\end{figure*}

\begin{figure*}[ht]
	\includegraphics[width=1.0\linewidth]{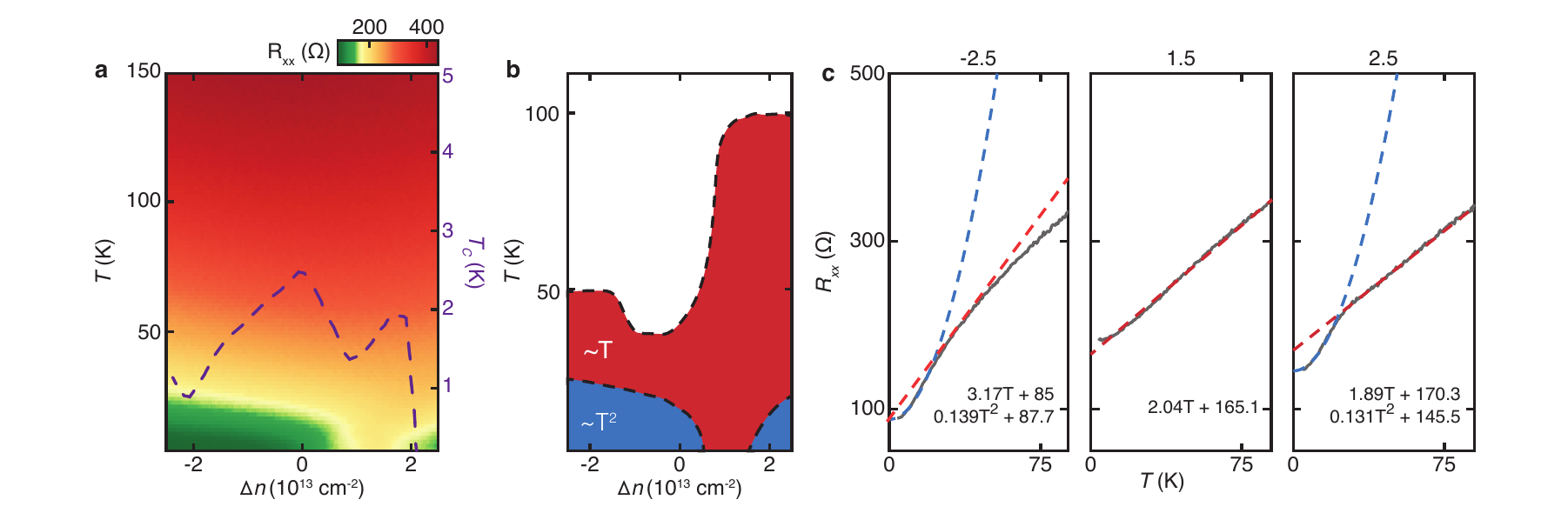}
	\caption*{\textbf{Extended Data Figure 2 $|$ Doping dependent T-linear behaviour of 2L $T_d$ - MoTe$_2$. a,} Colour plot of resistance vs temperature and doping. Doping dependent superconducting critical temperature (from Fig. 3a) is superimposed on the color plot with the associated temperature scale on the right axis. \textbf{b,} Diagram indicating regions of $T^2$ and $T$-linear behavior. \textbf{c,} $R_{xx}$ versus temperature for various dopings, as indicated at the top of each curve in values of $10^{13}$ cm$^{-2}$. Blue (red) dashed lines indicate quadratic (linear) fits to the data.}
\label{Ext2}
\end{figure*}

\begin{figure*}[ht]
	\includegraphics[width=1.0\linewidth]{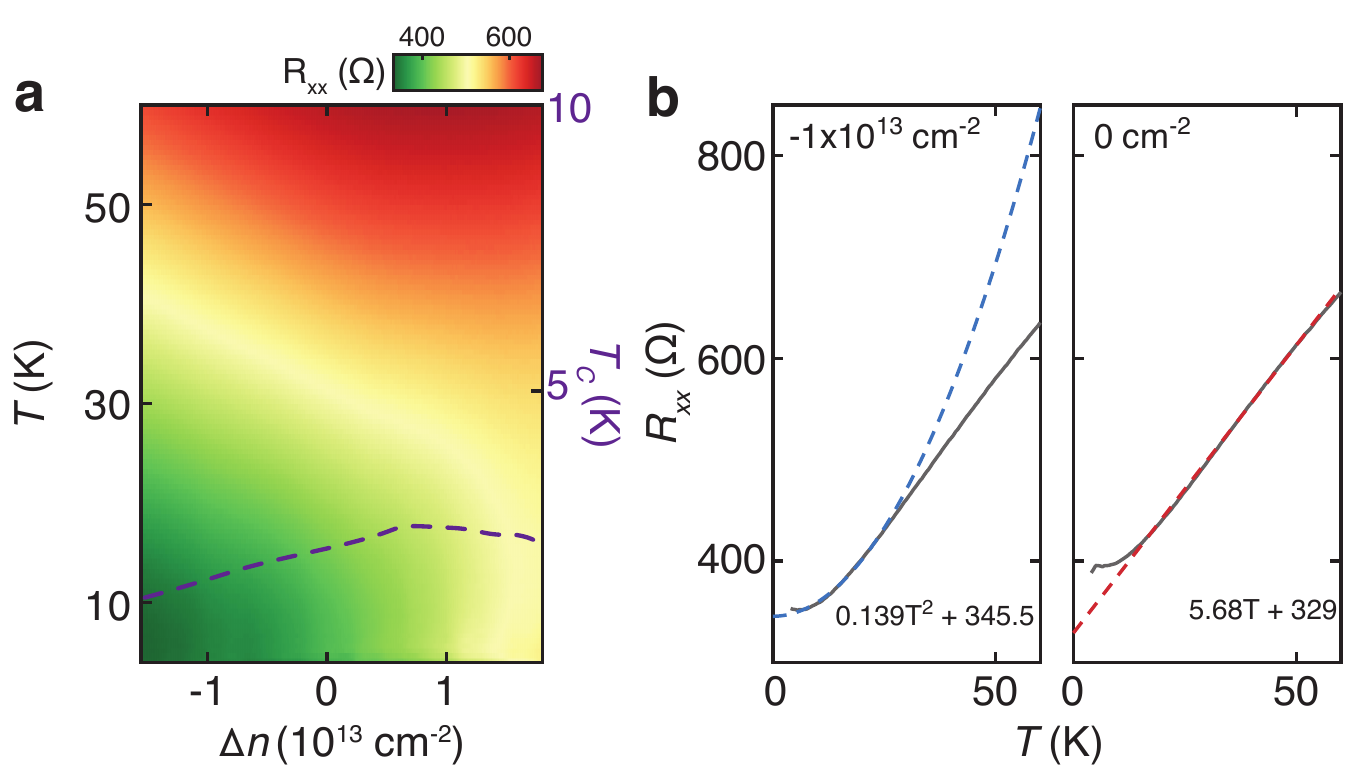}
	\caption*{\textbf{Extended Data Figure 3 $|$ T-linear behavior observed in D3. a,} Colour plot of resistance vs temperature and doping. Doping dependent superconducting critical temperature (from Extended Data Fig. 2) is superimposed on the color plot. \textbf{b,} $R_{xx}$ versus temperature for various dopings. Blue (red) dashed lines indicate quadratic (linear) fits to the data.}
\label{Ext3}
\end{figure*}

\clearpage
\begin{FlushLeft}
\large \textbf{Supplementary: Coupled Ferroelectricity and Superconductivity in Bilayer $T_d$-MoTe$_2$}
\end{FlushLeft}
\author{Apoorv Jindal$^{1}$, Amartyajyoti Saha$^{2,3}$, Zizhong Li$^{4}$, Takashi Taniguchi$^{5}$, Kenji Watanabe$^{5}$, James C. Hone$^{6}$, Turan Birol$^{3}$, Rafael M. Fernandes$^{2}$, Cory R. Dean$^{1}$, Abhay N. Pasupathy$^{1,\dagger}$, Daniel A. Rhodes$^{4,\dagger}$}

\maketitle
\begin{affiliations}
 \item Department of Physics, Columbia University, New York, NY, USA
 \item School of Physics and Astronomy, University of Minnesota, Minneapolis, MN, USA
  \item Department of Chemical Engineering and Materials Science, University of Minnesota, Minneapolis, MN, USA
 \item Department of Materials Science and Engineering, University of Wisconsin, Madison, WI, USA 
 \item National Institute for Materials Science, Tsukuba, Japan
 \item Department of Mechanical Engineering, Columbia University, New York, NY, USA
 \item[$^\dagger$] Correspondence to: apn2108@columbia.edu, darhodes@wisc.edu
\end{affiliations}
\clearpage
\section{Carrier density analysis}

Individual electron and hole carrier densities have been extracted by fitting the Hall resistance, R$_{xy}$, to a semiclassical two-band model \cite{MR_semimetals}. Since our magnetoresistance is sub-quadratic (Fig. 1d and Fig. S5), we are not able to simultaneously fit the longitudinal resistance, R$_{xx}$, to a model which inherently expects a quadratic behaviour. 
\[
\rho_{xx} (B) = \frac{(n\mu_n + p\mu_p) + (n\mu_p + p\mu_n)\mu_n\mu_pB^2}{e((n\mu_n + p\mu_p)^2 + (p - n)^2\mu_p^2\mu_n^2B^2)}
\]
\[
\rho_{xy}(B) = \frac{(p\mu_p^2 - n\mu_n^2)B + (p - n)\mu_p^2\mu_n^2B^3}{e((n\mu_n + p\mu_p)^2 + (p - n)^2\mu_p^2\mu_n^2B^2)}
\]
where $n(p)$ and $\mu_n(\mu_p)$ are electron (hole) density and mobility, respectively. We consider data for $|B| >$ 1 T to avoid fitting over the superconducting transition, where the Hall signal is comparatively weak. For ease of fitting, we place an upper bound for the electron mobility of 1250 cm$^{2}$/Vs which we extract from a single band fitting of the $\Delta n = 2.5 \times10^{13}\textrm{cm}^{-2}$ Hall resistance.

We also report a non-saturating subquadratic magnetoresistance (MR) (Fig. S5). Similar suppression of MR has been reported in other thin flake devices too \cite{low_MR_mote2,low_MR_wte2}. While this anomalous MR has been attributed to factors including reduced carrier mobility in thin flakes or enhanced substrate effects on charge transport, the origin of subquadratic MR remains to be investigated. The exponent drops in value as we move away from charge compensation by electrostatic doping as we move towards single carrier transport.

\clearpage

\begin{figure*}[ht] 
	\includegraphics[width=1\linewidth]{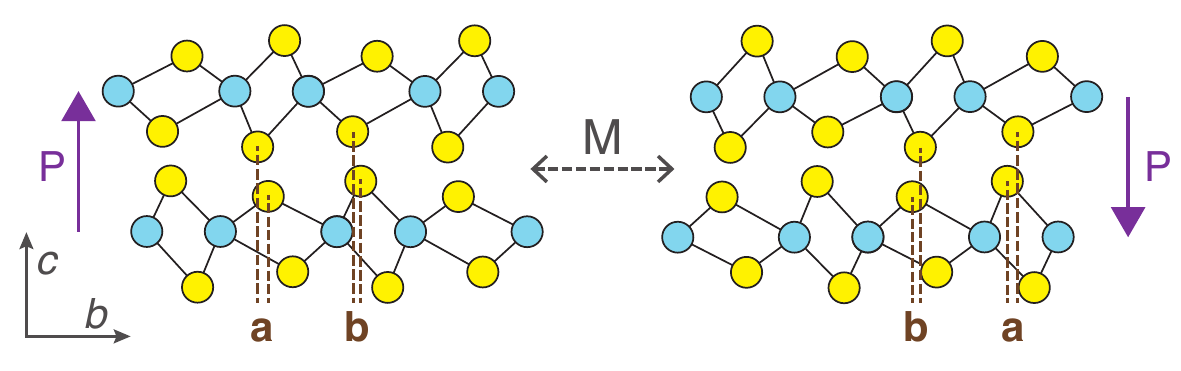}
	\caption*{\textbf{Fig. S1 $|$} Noncentrosymmetric degenerate ground states of 2L-MoTe$_2$. Interlayer sliding by $\Vec{a}+\Vec{b}$ is akin to a mirror operation along the horizontal axis leading to a switch in polarization\cite{wte2_ferro_reason, wte2_ferro_reason2}}
\label{Fig. 1}
\end{figure*}

\begin{figure*}[ht] 
	\includegraphics[width=1.0\linewidth]{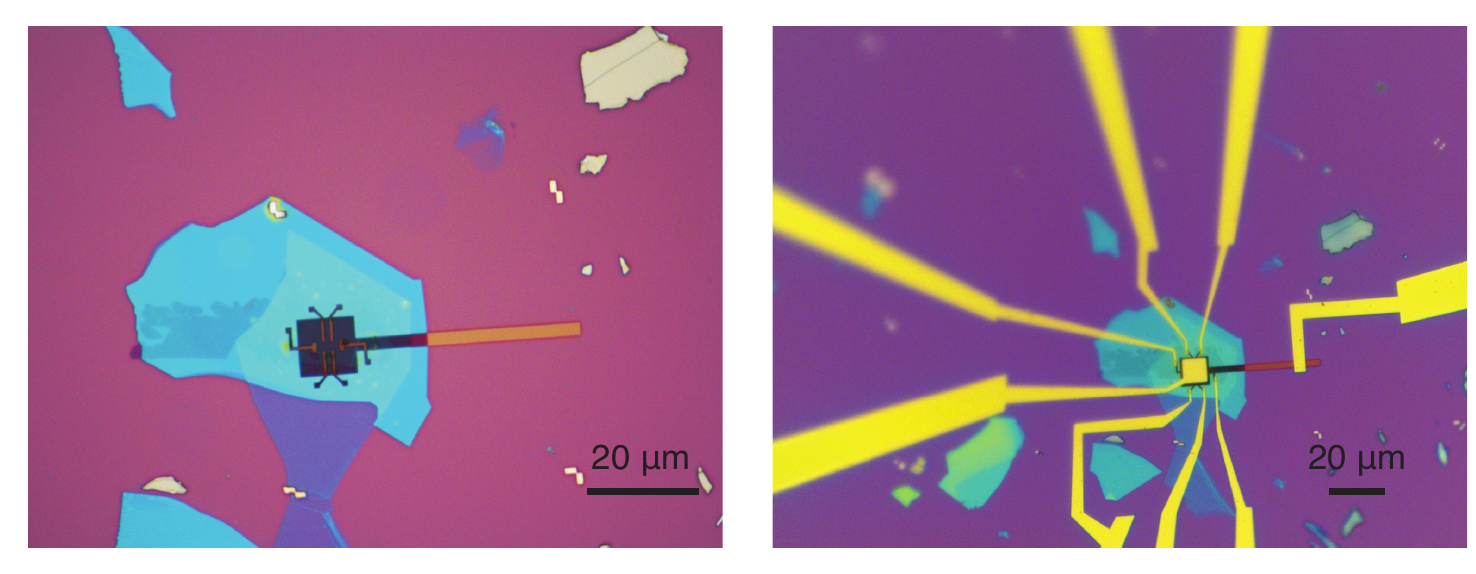}
	\caption*{\textbf{Fig. S2 Device image $|$} (a) after stacking, (b) after fabrication with metal leads and a metallic top gate. Top BN thickness used is 29 nm while the bottom BN is 7 nm thick.} 
\label{Fig. 2}
\end{figure*}

\begin{figure*}[ht] 
	\includegraphics[width=1.0\linewidth]{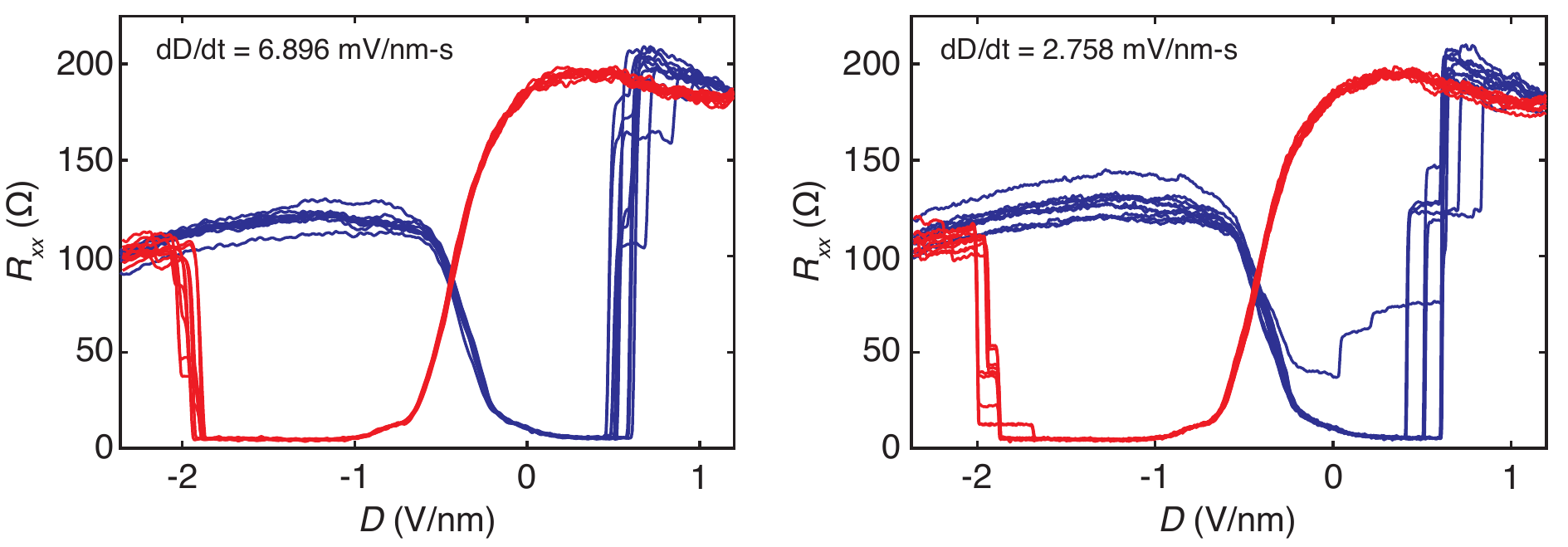}
	\caption*{\textbf{Fig. S3 Repeatability of ferroelectric switching $|$} Ferroelectric switching repeated over 10 cycles at varied sweep rates of displacement field. Blue (red) curves denote forward (backward) sweep directions in $D$.} 
\label{Fig. 3}
\end{figure*}

\begin{figure*}
	\includegraphics[width=1.0\linewidth]{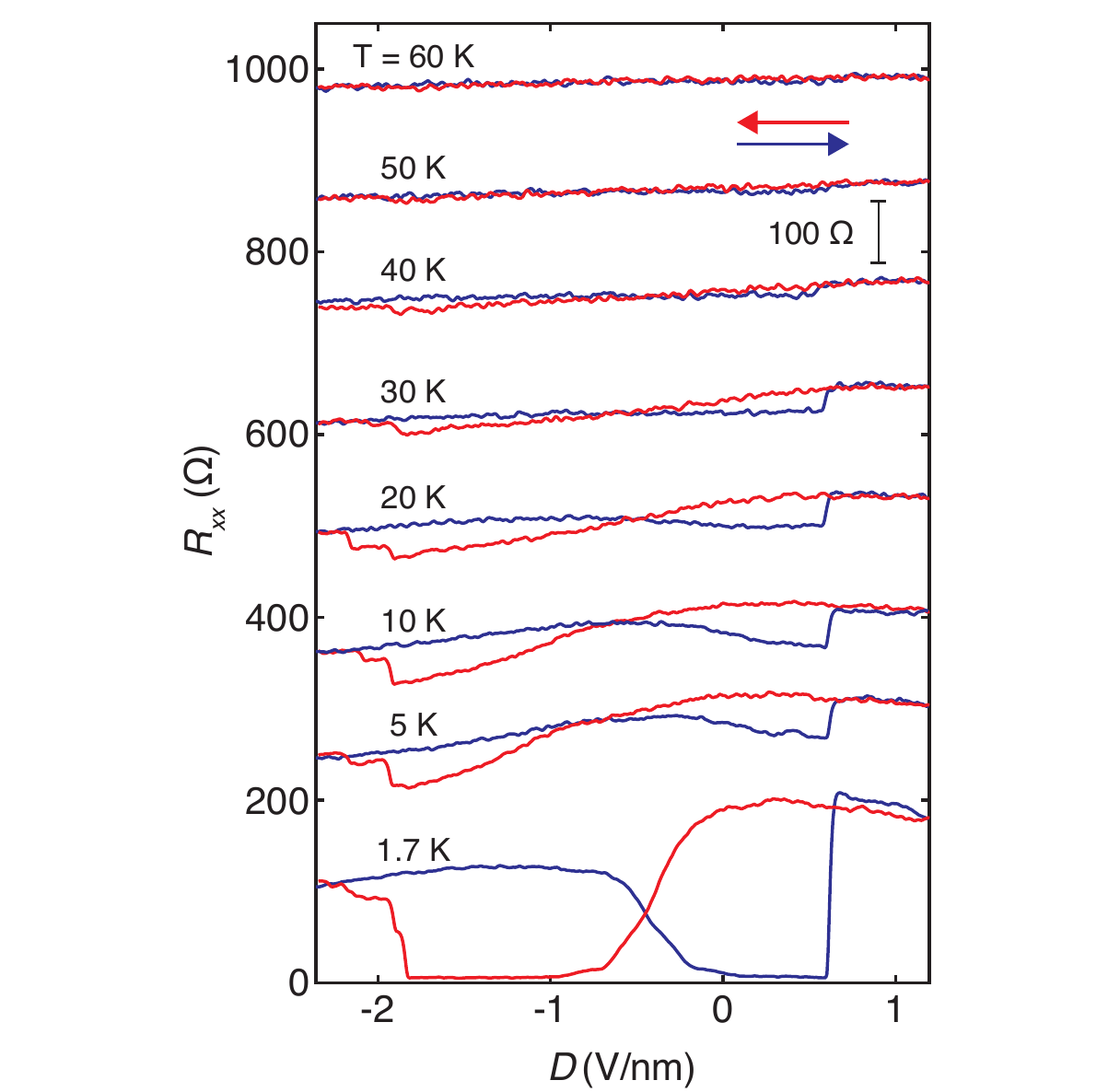}
	\caption*{\textbf{Fig. S4 Temperature dependence of ferroelectric switching $|$ } No switching observed above 50 K at $\Delta n$ = $1.5 \times 10^{13}$ cm$^{-2}$. The curves are offset by 100 $\Omega$ for clarity.} 
\label{Fig. 4}
\end{figure*}

\begin{figure*}
	\includegraphics[width=1\linewidth]{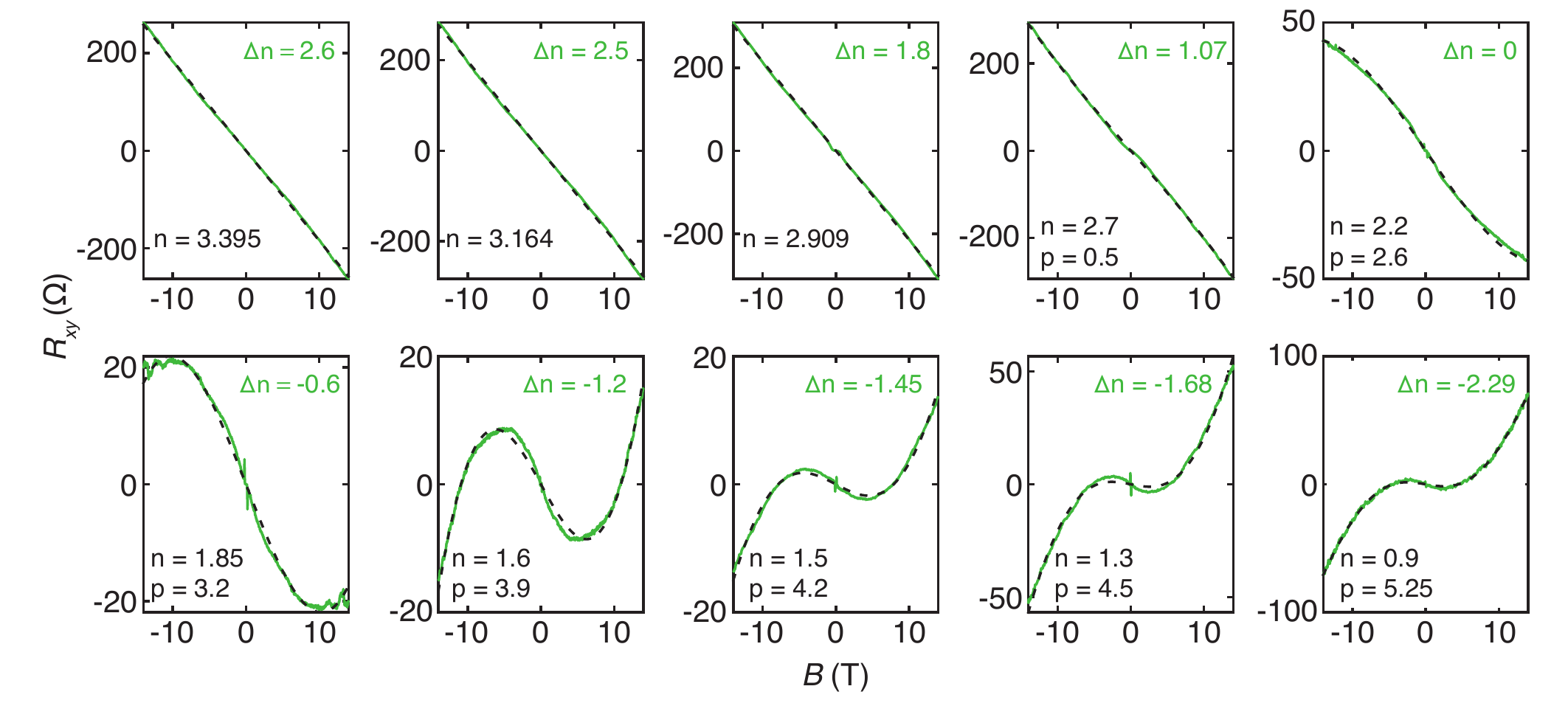}
	\caption*{\textbf{Fig. S5 $|$} Fitting two-carrier model to Hall data. Doping values indicated in each panel are in units of 10$^{13}$ cm$^{-2}$. Each panel corresponds to a specific value of $\Delta n$ (as defined in main text). All measurements were performed at 250 mK.}
\label{Fig. 5}
\end{figure*}

\begin{table}
\begin{center}
\caption{Extracted parameters from the two-band model. The first three entries were fit to a linear Hall response corresponding to $R_{xy}$ = B/ne with electron mobility calculated using the single band Drude formula, $\rho_{xx} = 1/ne\mu$.\vspace{0 mm}}
\begin{tabular}{ |c|c|c|c|c| } 
 \hline
 $\Delta n$ (10$^{13}$ cm$^{-2}$) & $n$ (10$^{13}$ cm$^{-2}$)  & $\mu_n$ (cm$^2$/Vs) & $p$ (10$^{13}$ cm$^{-2}$) & $\mu_p$ (cm$^2$/Vs)\\
 \hline
 2.6 & 3.395 & 1254 & - & -\\ 
 2.5 & 3.164 & 1253 & - & -\\ 
 1.8 & 2.909 & 1043 & - & -\\
 1.07 & 2.7 & 1156 & 0.5 & 559.3\\
 0 & 2.2 & 899.2 & 2.6 & 599.2\\
 -0.6 & 1.85 & 682.6 & 3.2 & 404.9\\
 -1.2 & 1.6 & 756.9 & 3.9 & 409.2\\
 -1.45 & 1.5 & 840.3 & 4.2 & 443.4\\
 -1.68 & 1.3 & 864.5 & 4.5 & 441\\
 -2.29 & 0.9 & 1000 & 5.25 & 404.9\\
 \hline
\end{tabular}
\label{table:1}
\end{center}
\end{table}
\clearpage
%

\begin{figure*}
	\includegraphics[width=1.0\linewidth]{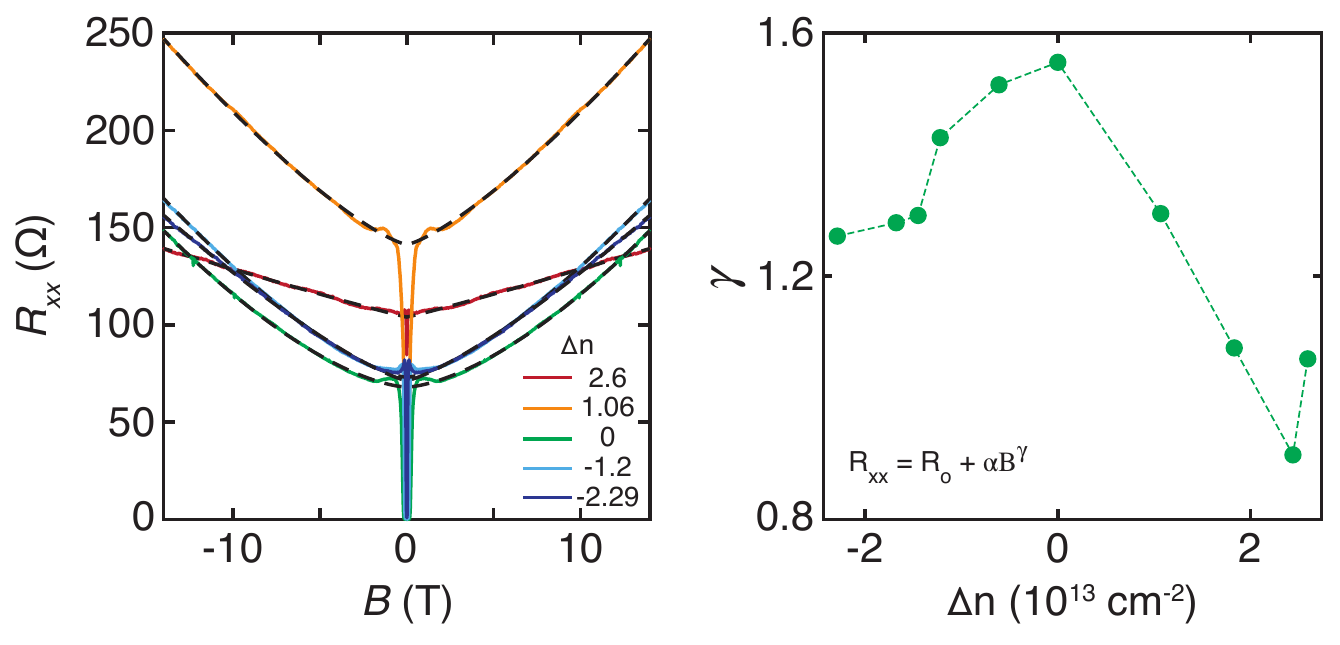}
	\caption*{\textbf{Fig. S6 $|$} Doping dependence of magnetoresistance, dashed lines are fits to the equation mentioned in the second panel.} 
\label{Fig. 6}
\end{figure*}

\begin{figure*}
	\includegraphics[width=1\linewidth]{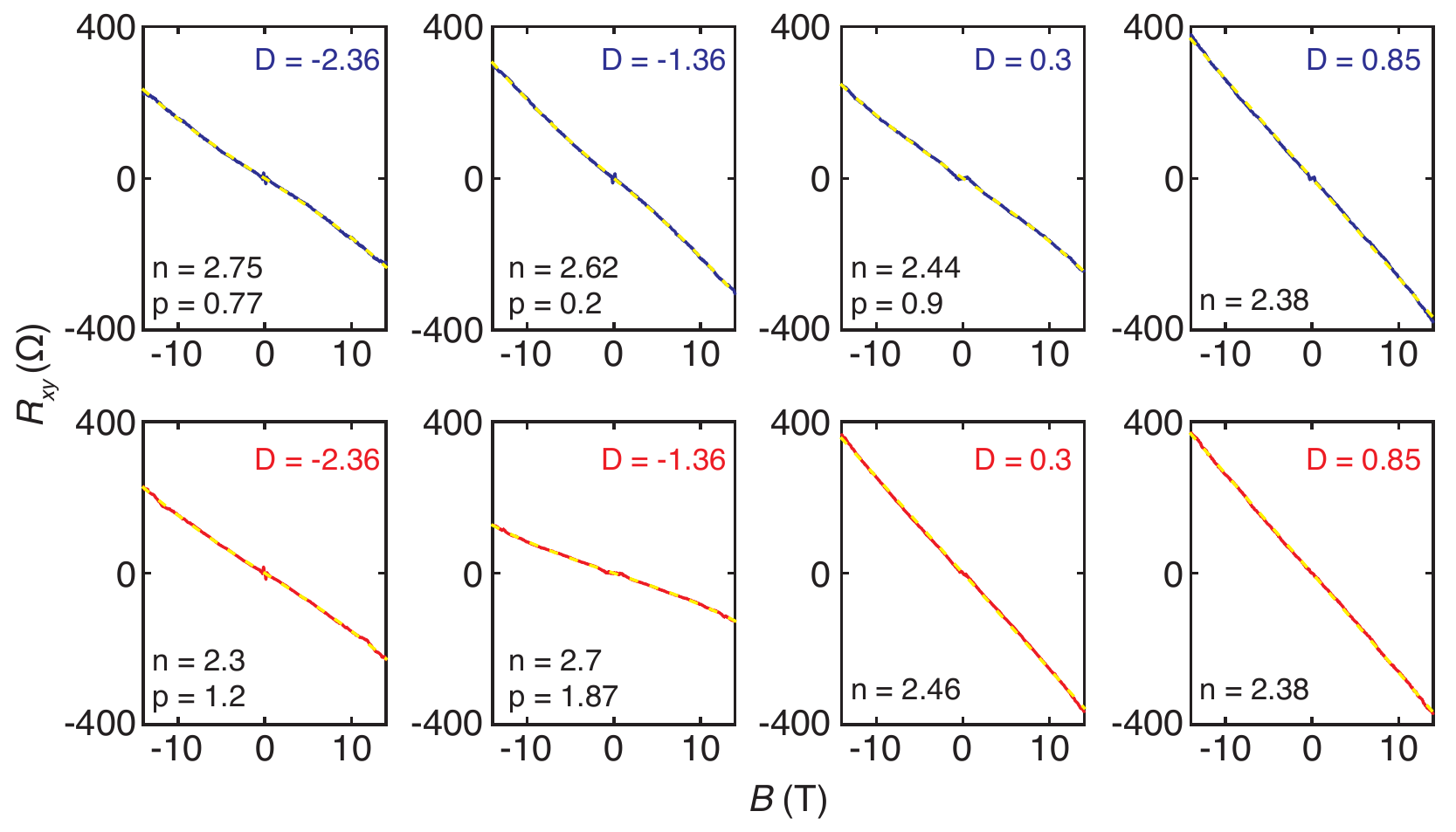}
	\caption*{\textbf{Fig. S7 $|$} Fitting two-carrier model to Hall data while sweeping displacement field across the butterfly loop in Fig. 2a of main text. Displacement field values indicated in each panel are in units of V/nm. Blue curves denote Hall data taken while $D$ was swept left to right in the butterfly loop while the red curves are taken while sweeping in the opposite direction. Dashed lines are fit to the two-band model and extracted doping values in each panel are in units of 10$^{13}$ cm$^{-2}$. All measurements were performed at 250 mK.}
\label{Fig. 7}
\end{figure*}

\begin{figure*}
\centering
	\includegraphics[width=0.65\textheight, keepaspectratio]{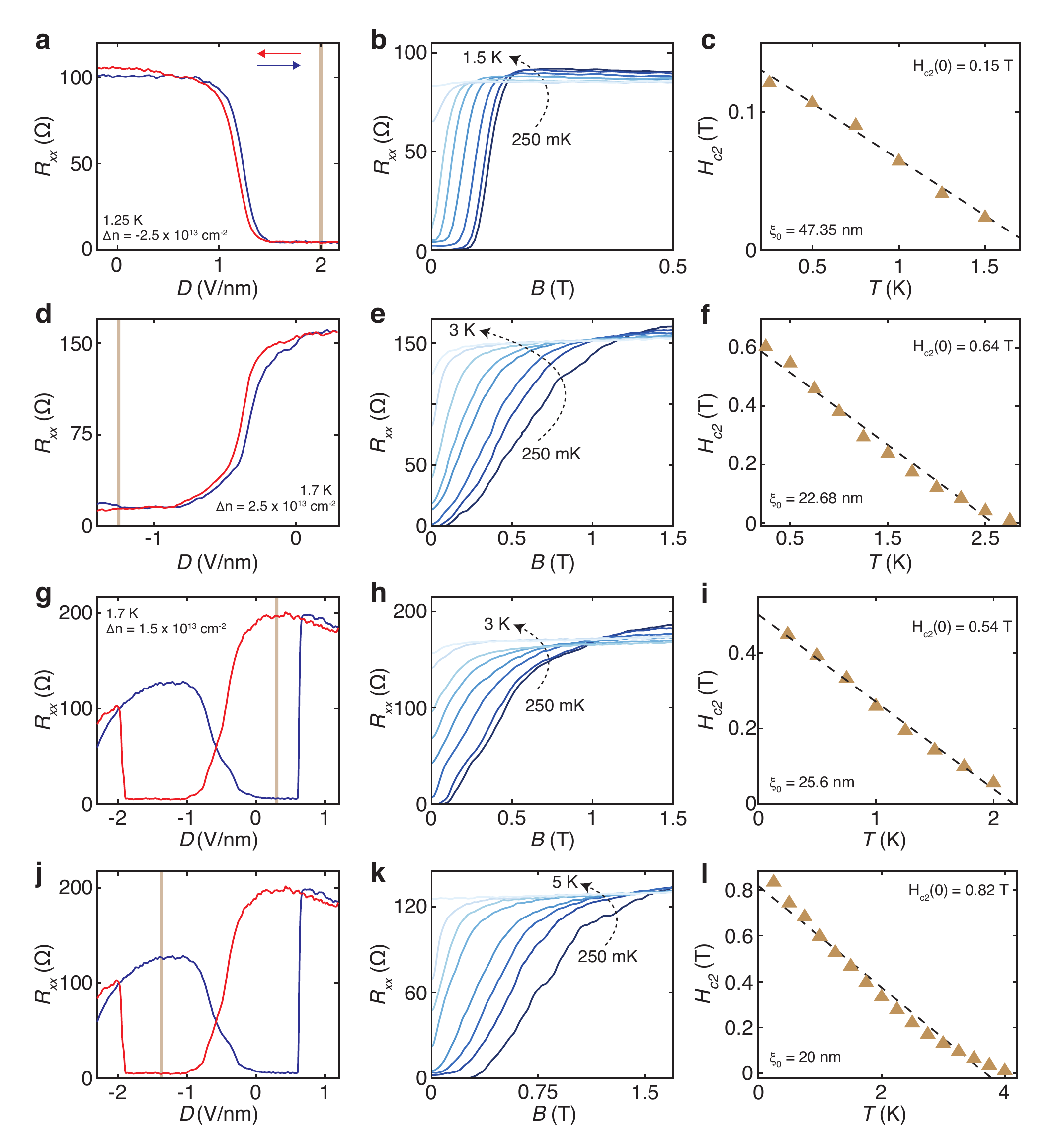}
	\caption*{\textbf{Fig. S8 $|$} B-T phase diagrams for field-induced and hysteretic superconductivity. \textbf{(a)}, Displacement field dependence of resistance at $\Delta n = -2.5\times10^{13}$/cm$^2$, highlighting the superconductivity at $D$ = 2 V/nm. \textbf={(b)}, Resistance vs field for different temperatures at the same $\Delta n$ and highlighted value of displacement field as indicated in \textbf{(a)}. \textbf{(c)}, $H^\perp_{c2}$ as a function of temperature as extracted from \textbf{(b)} using $R_{xx}$ = 50\% of $R_\textrm{n}$. The dashed line is a fit to the 2D GL model for out-of-plane magnetic fields. \textbf{(d-l)}, The same as \textbf{(a-c)}, but for different values of $\Delta n$ and $D$.} 
\label{Fig. 8}
\end{figure*}

\clearpage

\end{document}